%% file: main.tex
\documentclass{article}

\usepackage{microtype}
\usepackage{graphicx}
\usepackage{subfigure}
\usepackage{booktabs} 
\usepackage{url}
\usepackage{amsmath}
\usepackage{amssymb}
\usepackage{amsfonts}
\usepackage{amsthm}
\usepackage{authblk}
\usepackage{courier}

\usepackage{hyperref}

\usepackage[accepted]{icml2020}

\icmltitlerunning{ProGen: Language Modeling for Protein Generation}

\begin{document}

\twocolumn[
\icmltitle{ProGen: Language Modeling for Protein Generation}



\icmlsetsymbol{equal}{*}

\begin{icmlauthorlist}
\icmlauthor{Ali Madani}{salesforce}
\icmlauthor{Bryan McCann}{salesforce}
\icmlauthor{Nikhil Naik}{salesforce}
\icmlauthor{Nitish Shirish Keskar}{salesforce}
\icmlauthor{Namrata Anand}{stanford}
\icmlauthor{Raphael R. Eguchi}{stanford}
\icmlauthor{Po-Ssu Huang}{stanford}
\icmlauthor{Richard Socher}{salesforce}
\end{icmlauthorlist}

\icmlaffiliation{salesforce}{Salesforce Research, Palo Alto, CA, USA}
\icmlaffiliation{stanford}{Department of Bioengineering, Stanford University, Stanford, CA, USA}

\icmlcorrespondingauthor{Ali Madani}{amadani@salesforce.com}

\icmlkeywords{language modeling, generative, protein, sequences, controllable, conditional}

\vskip 0.3in
]



\printAffiliationsAndNotice{} 

\input{section/0-abstract}
\input{section/1-introduction}
\input{section/2-related-works}
\input{section/3-methods}
\input{section/4-results}

\input{section/5-conclusion}
\input{section/6-acknowledgements}
\clearpage
\bibliography{section/7-references}
\bibliographystyle{icml2020}
\clearpage
\appendix
\input{section/8-appendix}
\end{document}

%% file: section/0-abstract.tex
\begin{abstract}
Generative modeling for protein engineering is key to solving fundamental problems in synthetic biology, medicine, and material science.
We pose protein engineering as an unsupervised sequence generation problem in order to leverage the exponentially growing set of proteins that lack costly, structural annotations.
We train a 1.2B-parameter language model, ProGen, on $\sim$280M protein sequences conditioned on taxonomic and keyword tags such as molecular function and cellular component.
This provides ProGen with an unprecedented range of evolutionary sequence diversity and allows it to generate with fine-grained control as demonstrated by metrics based on primary sequence similarity, secondary structure accuracy, and conformational energy.
\end{abstract}

%% file: section/1-introduction.tex
\section{Introduction}
\label{introduction}
\begin{figure*}[h]
\vskip 0.2in
\begin{center}
\centerline{\includegraphics[width=\linewidth]{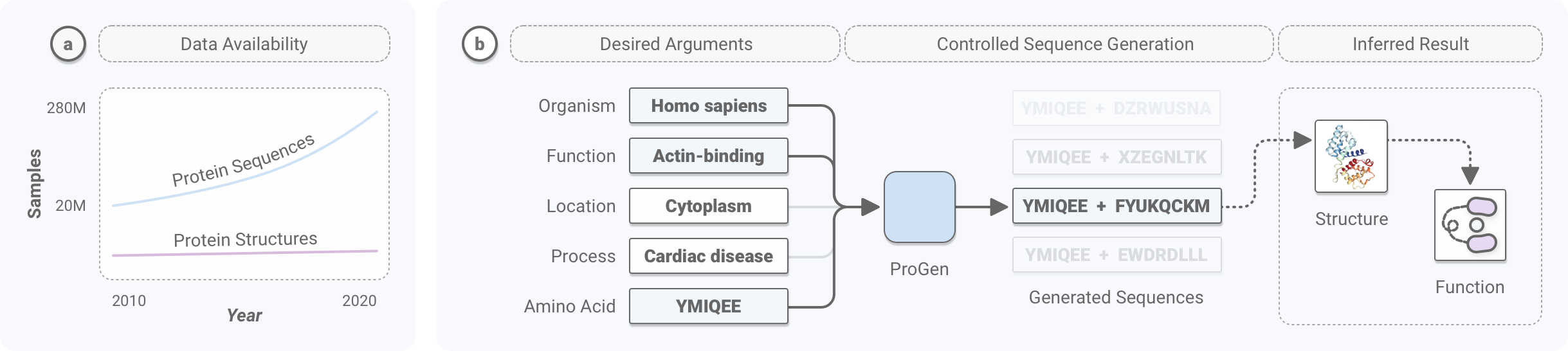}}
\caption{a) Protein sequence data is growing exponentially as compared to structural data. b) We utilize protein sequence data along with taxonomic and keyword tags to develop a conditional language model: ProGen.}
\label{intro_fig}
\end{center}
\vskip -0.2in
\end{figure*}

Generating proteins with desired properties is one of the most complex yet impactful problems in biology. 
Protein engineering research has grown over the past 50 years and yielded remarkable outcomes including the development of new enzymes, therapies, and sensors.
However, leading experimental techniques for protein engineering such as directed evolution~\citep{arnold1998design} still rely on heuristics and random mutations to select initial sequences for rounds of evolution.

The raw amino acid sequence encodes a protein, and during synthesis, this chain of amino acids folds in ways that exhibit local (secondary) and global (tertiary) structure.
These structural properties then directly determine a unique function, which is of ultimate interest to protein engineers.
Unfortunately, obtaining three-dimensional structural information for proteins is expensive and time consuming. Consequently, there are three orders of magnitude more raw sequences than there are sequences with structural annotations, and protein sequence data is growing at a near exponential rate.

Recent research~\citep{alley2019unified,rives2019biological,rao2019evaluating} has begun to capitalize on the much larger set of raw protein sequences by adapting state-of-the-art representation learning techniques~\citep{devlin2018bert} from natural language processing (NLP) to classification of protein properties.
However, there has been no attempt to adapt state-of-the-art methods for artificial text generation~\citep{radford2019language}, and in particular the kind of controllable generation~\citep{keskar2019ctrl} that would be most useful for protein engineering.

We introduce ProGen for controllable protein generation.
ProGen is a 1.2 billion parameter conditional language model trained on a dataset of 280 million protein sequences together with conditioning tags that encode a variety of annotation such as taxonomic, functional, and locational information.
By conditioning on these tags, ProGen provides a new method for protein generation that can be tailored for desired properties (Figure~\ref{intro_fig}).

According to NLP metrics, ProGen is a powerful language model, achieving comparable performance to similarly-sized models for English.
This performance improves in settings with larger amino acid contexts and when ProGen is provided a larger number of conditioning tags, which highlights its potential for applications in providing viable, starting sequences for directed evolution or \textit{de novo} protein design~\citep{huang2016coming}.
ProGen also performs well when used to model unseen protein families, but it is even more effective when fine-tuned for those unseen families as an alternative to training from random initialization.
These results inspire the use of ProGen to generate candidate sequences in challenging, low-homology applications.

Proteins generated by ProGen satisfy desired structural, and by extension functional, properties when evaluated with metrics for sequence similarity, secondary structure accuracy, and conformational energy-- from lower level structure to higher level structure.
Generation performance is judged higher quality by higher level metrics, which suggests that ProGen has learned invariances to mutations at the sequence level that conserve structure and inferred function.
At the highest level, conformational energy analysis reveals that generated proteins exhibit energy levels near that of native proteins, providing our strongest evidence that these proteins satisfy the desired structure and inferred function.
In our first case study, we examine completion of a VEGFR2 protein, which is held-out in training.
ProGen generates candidate completions that conserve the structural elements most important for determining function and exhibit conformational energy near to native levels across a variety of generation lengths. 
In our second case study, we observe that ProGen can select high fitness antibody-binding GB1 proteins without supervised training or unsupervised finetuning-- indicating that ProGen has learned the underlying distribution of functional proteins.

%% file: section/2-related-works.tex
\section{Related Work}
\label{sec:related_work}
\textbf{Protein representation learning.}
Recent methods for contextualized representations~\citep{mccann2017learned,peters2018deep,devlin2018bert}  in natural language processing have been demonstrated to work well for contextual protein representation learning. 
Structural information about a protein can be extracted from such representations using linear methods, and the representations themselves can be adapted to improve performance on other tasks~\citep{rives2019biological}.
Similarly, UniRep~\citep{alley2019unified} demonstrated that such representations could be used to predict stability of natural and {\it de novo} designed proteins as well as quantitative function of molecularly diverse mutants.
TAPE~\citep{rao2019evaluating} is a new benchmark consisting of five tasks for assessing such protein embeddings.
While this body of prior work focuses on transferable representation learning using bidirectional models, our work demonstrates controllable protein engineering with generative, unidirectional models.

\textbf{Generative models for protein engineering.}
Recent generative modeling work such as ~\citet{ingraham2019generative} extends the transformer to condition it on a graph-structured specification of a desired target.~\citet{anand2018generative} utilizes generative adversarial networks to produce 2D pairwise distance map for given protein structural fragments, essentially in-painting missing residues. The aforementioned work, along with ~\citet{o2018spin2}, ~\citet{boomsma2017spherical}, and ~\citet{greener2018design}, all utilize explicit structural information for generative modeling, thereby are unable to fully capture the number and diversity of sequence-only data available.
Meanwhile sequence-only generative modeling have been attempted recently through residual causal dilated convolutional neural networks ~\citep{riesselman2019accelerating} and variational autoencoders ~\citep{costello2019hallucinate}.
Unlike these prior works, our work on generative modeling focuses on a high-capacity language models that scale well with sequence data and can be used for controllable generation.

\textbf{Language Models and Controllable Generation.}
 Large Transformer architectures~\citep{vaswani2017attention} like GPT-2~\citep{radford2019language} represent the state-of-the-art in unconditional language modeling and demonstrate impressive text generation capabilities~\citep{zellers2019defending} after training on vast amounts of unsupervised English text. 
 CTRL~\citep{keskar2019ctrl} trained a similarly large Transformer architecture for language generation by conditioning on properties of the text easily extracted at scale, e.g. domain, style, and even associated URL. 
 We adapt this perspective to protein engineering by training a conditional transformer language model on amino acid sequences conditioned on a set of protein properties referred to as conditioning tags. 
 Notably different from~\citet{keskar2019ctrl}, protein engineering requires a finer-grained, much larger, and more complex set of conditioning tags. 
 Additionally, a single protein can be paired with dozens of conditioning tags.

%% file: section/3-methods.tex
\section{Methods}
\label{sec:methods}
Let $a=(a_1, \ldots, a_{n_a})$ be a sequence of amino acids that constitutes a protein.
In the context of protein engineering, there is typically also a set of desired protein properties such as function or affiliation with a particular organism.
Following recent work on controllable, conditional language modeling~\citep{keskar2019ctrl}, we refer to these properties generally as `conditioning tags' through which we would like to control generation of amino acid sequences. Let $c=(c_1, \ldots, c_{n_c})$ be a sequence of such conditioning tags, and let $x=[c;a]$ the sequence formed by prepending a conditioning tag sequence to an amino acid sequence. $p(x)$ is then the probability over such combined sequences of length $n=n_a + n_c$. We can factorize this distribution using the chain rule of probability~\citep{bengio2003neural}:
\[ p(x) = \prod_{i=1}^n p(x_i | x_{<i}) \]
This decomposes the problem of conditional protein generation into next-token prediction, where a token $x_i$ can either be an amino acid or a conditioning tag.
A neural network with parameters $\theta$ can then be trained to minimize the negative log-likelihood over a dataset $D=\{x^{1}, \ldots, x^{|D|}\}$: 
\[ \mathcal{L}(D) = - \sum_{k=1}^{|D|} \log {p_{\theta}(x_{i}^{k} | x^{k}_{<i})} \]
Note that $p(a|c)$, the distribution over proteins conditioned on their corresponding conditioning tags, is just one of the many conditional distributions that can be recovered from a model that learns $p(x)$. 
A new protein $\tilde a$ of length $m_a$ with desired properties encoded by a conditioning tag sequence $\tilde c$ of length $m_c$  can then be generated by sequentially sampling its constituent symbols: 
$p_{\theta}(a_0|\tilde c), p_{\theta} (a_1 | \tilde a_0,\tilde c), \dots, p_{\theta}(a_p | \tilde a_{<p},\tilde c)$.

We train a variant of the Transformer~\citep{vaswani2017attention} to learn these conditional distributions over amino acids and conditioning tags.
A sequence containing $n$ tokens is embedded as a sequence of $n$ corresponding vectors in $\mathbb{R}^d$.
Each vector is the sum of a learned token embedding and a sinusoidal positional embedding as in the original Transformer architecture.
This sequence of vectors is stacked into a matrix $X_0\in\mathbb{R}^{n \times d}$ so that it can be processed by $l$ attention layers.
The $i$th layer consists of two blocks, each of which preserves the model dimension $d$.

The core of the first block is multi-head attention with $k$ heads that uses a causal mask to preclude attending to future tokens:
\begin{align}
\text{Attention}(X, Y, Z)&=\text{softmax}\left(\frac{\text{mask}(XY^\top)}{\sqrt{d}}\right) Z\nonumber\\
\text{MultiHead}(X, k) &= [h_1;\cdots;h_k]W_o\nonumber\\
\text{where } h_j &= \text{Attention}( XW_j^1,  XW_j^2,  XW_j^3) \nonumber
\end{align}

The core of the second block is a feedforward network with ReLU activation that projects inputs to an inner dimension $f$, with parameters $U\in \mathbb{R}^{d \times f}$ and $V\in \mathbb{R}^{f \times d}$:
\begin{align}
FF(X) = \text{max}(0, XU)V \nonumber
\end{align}

Each block precedes core functionality with layer normalization~\citep{Ba2016LayerN, child2019sparse} and follows it with a residual connection~\citep{he2016deep}. Together, they yield $X_{i+1}$:
\begin{align}
 \text{\underline{Block 1}} &&& \text{\underline{Block 2}} \nonumber
\end{align}
\begin{align}
\bar X_i &= \text{LayerNorm}(X_i) & \bar H_i &= \text{LayerNorm}(H_i) \nonumber\\
 H_{i} &= \text{MultiHead}(\bar X_i) + \bar X_i &  X_{i+1} &= \text{FF}(\bar H_i) + \bar H_i\nonumber 
\end{align}

Scores are then computed from the output of the last layer:
\[ \text{Scores}(X_0) = \text{LayerNorm}(X_{l}) W_{vocab} \]
During training, these scores are the inputs of a cross-entropy loss function.
During generation, the scores corresponding to the final token are normalized with a softmax, yielding a distribution for sampling a new token.

\subsection{Data}
We utilize all protein sequences and associated tags available in Uniparc~\citep{leinonen2004uniprot}, UniprotKB~\citep{bairoch2005universal}, SWISS-PROT~\citep{bairoch2004swiss}, TrEMBL~\citep{boeckmann2003swiss}, Pfam~\citep{bateman2004pfam}, and NCBI taxonomic information~\citep{federhen2012ncbi}.
The aggregated dataset contains over $281$M proteins---the most comprehensive, non-redundant, annotated database of proteins used to train a machine learning model. 
For the amino acid vocabulary, we use the standard 25 amino acids designations in IUPAC ~\citep{pettit2006iupac}.
The conditioning tags are divided into 2 categories: (1) keyword tags and (2) taxonomic tags. Following the definitions laid out in the UniprotKB controlled, hierarchical vocabulary of keywords (many of which are derived from Gene Ontology (GO) terms)~\citep{ashburner2000gene}, the conditioning keyword tags included 1100 terms ranging from cellular component, biological process, and molecular function terms. The taxonomic tags include 100k terms from the NCBI taxonomy across the eight standard taxonomic ranks. 
The aggregated dataset was split into a training set of size 280M, a held-out protein family\footnote{Protein families are groups of evolutionarily-related proteins that have similar structure, function, and sequence similarity as defined by Pfam~\citep{bateman2004pfam}} test set (OOD-test) of size 100k, and a randomly sampled test set (ID-test) of size 1M.
OOD-test comprises of 20 protein families, as defined in Pfam, that were excluded from the training data. 
Performance on OOD-test measures ability to model samples from unseen protein families, whereas performance on ID-test measures ability to model samples from a wider range of protein families that more closely match the distribution of the training set as described in section~\ref{sec:OOD}

\subsection{Training Details}
For training, we include each sequence and its reverse, as proteins are invariant to the temporal notion of sequence generation. 
We then prepend each sequence (and its reverse) with a corresponding subset of conditioning tags.
For a given sequence, there can be multiple versions across databases, each with their own associated conditioning tags. In training, we randomly sample which set of conditioning tags to utilize but bias toward SWISSPROT tags as they are manually verified. We apply dropout to the conditioning tags themselves at a rate of $0.4$.
We additionally always include a sample with the sequence alone without conditioning tags so that ProGen can be used to complete proteins using only sequence data even when no protein properties are known.
We then truncate all sequences to a maximum length of $512$.
Sequences of length less than $512$ were padded, but no loss was backpropagated through the network for padding tokens. The model has dimension $d=1028$, inner dimension $f=512$, $36$ layers, and $8$ heads per layer. Dropout with probability $0.1$ follows the residual connections in each layer. Token embeddings were tied with the embeddings of the final output layer~\citep{inan,press2016using}. 

Our model was implemented in TensorFlow~\citep{tensorflow} and trained with a global batch size of $64$ distributed across $256$ cores of a Cloud TPU v$3$ Pod for $1$M iterations. 
Training took approximately two weeks using Adagrad~\citep{duchi2011adaptive} with linear warmup from $0$ to $1e^{-2}$ over $40$k steps.
Gradient norms were clipped to $0.25$. Training in early stages was improved and stabilized by initializing with the pretrained weights of~\citet{keskar2019ctrl}. 

\subsection{Generation Details}
\label{subsec:generation_details}
ProGen generates proteins one amino acid at a time.
For one step of generation, ProGen takes  a context sequence of amino acids as input and outputs a probability distribution over amino acids. We sample from that distribution and then update the context sequence with the sampled amino acid.
This process repeats until a protein of desired length has been generated.
We compare different combinations of top-$k$ sampling~\citep{radford2019language} with a repetition penalty designed for amino acid sequence generation.  The repetition penalty reduces the probability of amino acids that have been generated within $4$ tokens prior to the token to be predicted.
Top-$k$ sampling draws the next token from the $k$ most probable tokens in the distribution output by ProGen.
We report results for top-$k$ values of $k=1$ and $k=3$ with repetition penalties of $0$ and $1.2$.

\subsection{Evaluation Details}
\label{subsec:evaluation_details}
To assess how well ProGen models the training and test distributions, we rely on perplexity as the standard metric for language models, a mean hard accuracy over each token to strictly assess each amino acid error, and a mean soft accuracy defined by incorporating BLOSUM62~\citep{henikoff1992amino}, a standard amino acid substitution matrix.

Perplexity is the exponentiated cross-entropy loss computed over each token in a dataset.
Thus, high quality language models are expected to have low perplexities. 
Mean per-token hard accuracy over the tokens in a sequence judges a prediction incorrect for any amino acid that is not the ground truth.
Mean per-token soft accuracy relies on BLOSUM62, a block substitution matrix that specifies which amino acid substitutions are more or less acceptable according to their frequency in known well-formed proteins. BLOSUM62 is widely used across adopted alignment software (e.g., BLAST\footnote{https://blast.ncbi.nlm.nih.gov/Blast.cgi}).
Our mean per-token soft accuracy uses BLOSUM62 to penalize incorrect amino acid predictions according to the frequency of that substitution in the matrix.
In this way, if the substitution is likely in nature, soft accuracy penalizes the model less.

To assess the quality of generation, we evaluate across three levels of structure: (1) primary sequence similarity, (2) secondary structure accuracy, and (3) conformational energy analysis.

Primary sequence similarity is defined by a global, pairwise sequence alignment score computed with the Biopython package\footnote{https://biopython.org/}. 
This score is based on the Needleman-Wunsch algorithm~\citep{needleman1970general} informed by the BLOSUM62 substitution matrix.
We use a gap open penalty of $-0.5$ and gap continue penalty of $-0.1$. 
The resulting score is then normalized by the length of the protein. 
Experiments reporting sequence similarity are limited to test samples with a form of experimental evidence of X-ray/NMR crystallography, mass spectrometry, or existence in cDNA or RT-PCR to indicate transcript existence. We refer the reader to UniprotKB existence scores with experimental evidence\footnote{https://www.uniprot.org/help/protein\_existence} for further details.

Secondary structure accuracy was computed per-residue for predicted secondary structures by PSIPRED\footnote{http://bioinf.cs.ucl.ac.uk/psipred/} with greater than 0.5 confidence. 
PSI-BLAST was performed on each generated sample to extract the Multiple Sequence Alignments (MSAs) with respect to the UniRef90 database~\citep{suzek2015uniref}. 
These MSAs were provided to PSIPRED for higher quality secondary structure prediction.
Experiments reporting secondary structure accuracy were limited to test samples with high UniprotKB existence scores as described in the previous paragraph.

Conformational energy uses the Rosetta-RelaxBB protocol\footnote{https://www.rosettacommons.org/}. 
Rosetta-RelaxBB performs a Monte Carlo optimization of the Rosetta energy function over the space of amino acid types and rotamers. 
The Rosetta energy is based on biophysical laws and constraints. 
Between each design round, amino acid side-chains are replaced, while the carbon backbone torsions are kept fixed. 
Energy minimization/relaxation is performed after threading the amino acid sequence through the known structure.
This allows the backbone to move, possibly into a lower energy state. 
A lower resulting Rosetta energy correlates to a more relaxed-state and viable conformation for a given protein structure.
Before applying the procedure above, we relax the native template first.
Experiments that report conformational energy are limited to test samples from SWISSPROT with associated 3D structures in RCSB PDB~\footnote{https://www.rcsb.org/}.

\begin{table}[t]
\caption{ ProGen outperforms uniform random and empirical baselines on the full test set, which includes ID- and OOD-test. OOD-test results reveal that ProGen also performs well on protein families unseen during training. Fine-tuning ProGen dramatically improves performance over training from random initialization. }
\label{test_set_ppl_acc}
\vskip 0.15in
\begin{center}
\begin{small}
\begin{sc}
\begin{tabular}{lccr}
\toprule
Model                   & PPL & Hard Acc. \\
\midrule
Uniform Baseline  & 25         & 4    \\
Empirical Baseline & 18.14     & 6    \\
ProGen                  & 8.56      & 45   \\
\midrule
\hspace{3mm}ID-test & 8.17 & 45 \\
\hspace{3mm}OOD-test & 13.34 & 22 \\
\midrule
\hspace{3mm}OOD-test-20 (rand. init.) & 17.78 & 9 \\
\hspace{3mm}OOD-test-20 (fine-tuned) & 7.45 & 50 \\
\bottomrule
\end{tabular}
\end{sc}
\end{small}
\end{center}
\end{table}

To assess generative quality, we provide baselines for different levels of random mutation.
For a given sequence, a proportion ($25-100\%$) of amino acids in the sequence is randomly substituted within one of the $20$ standard amino acids other than itself.
For conformational energy, we also include an all-alanine baseline (i.e. a sequence with only the amino acid alanine), as it is a non-bulky, chemically inert amino acid that mimics the existing secondary structure well when substituted. 
These baselines provide a scale across each of the above metrics.
A particular random mutation may or may not have constructive or destructive effects on protein structure or function.
But viewed in aggregate, the performance of the $100\%$ mutation baseline for any metric indicates failed generation.
As performance approaches $0\%$, generation statistically indicates a closer reflection to desired structural and functional properties.

\begin{figure}[t]
\vskip 0.2in
\begin{center}
\centerline{\includegraphics[width=1\columnwidth]{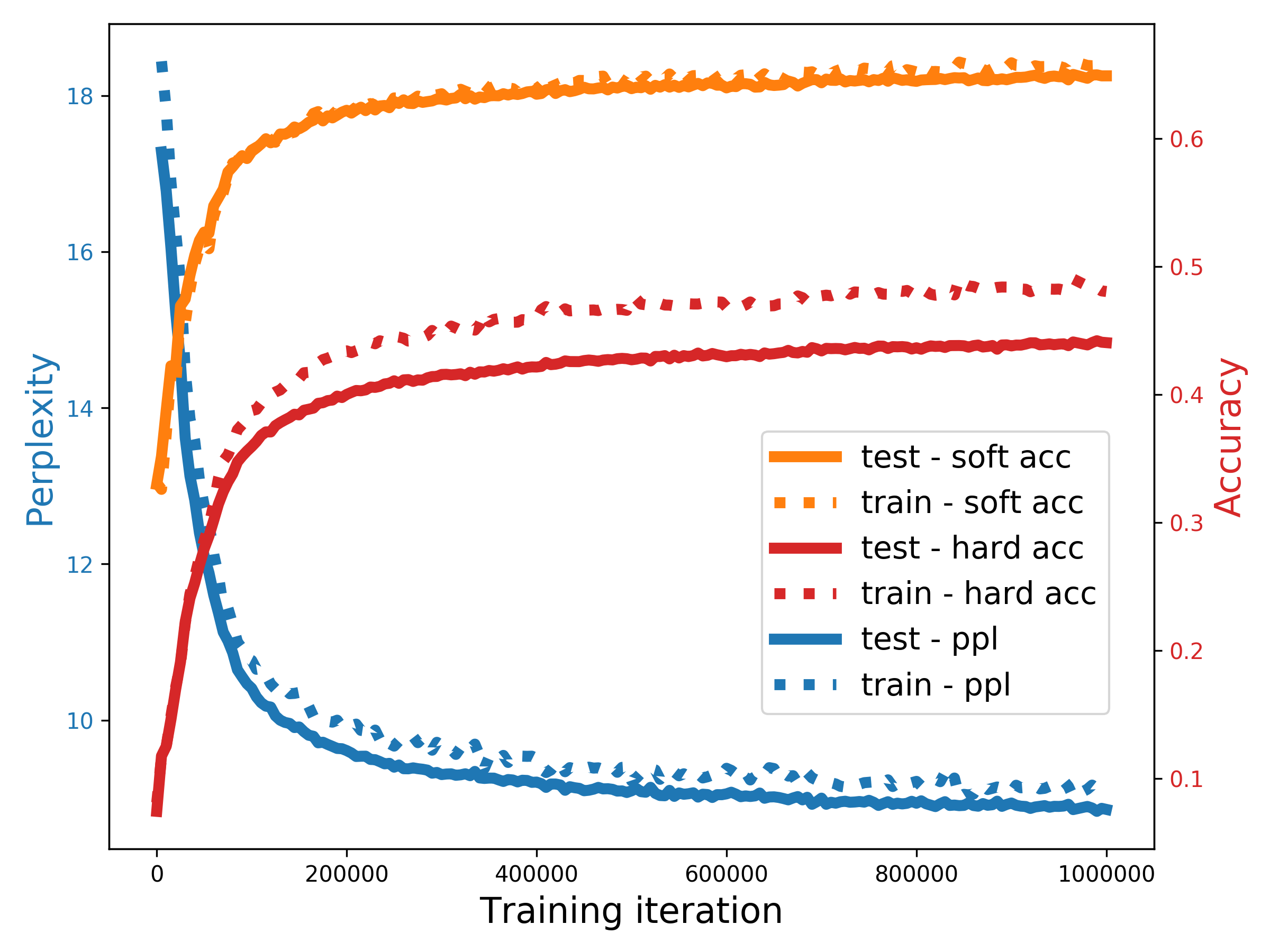}}
\caption{Large model capacity is warranted as ProGen has yet to overfit. BLOSUM62-informed soft accuracy shows no gap between train and test performance, suggesting hard accuracy hides the possibility that ProGen errors often correspond to amino acid substitutions found in nature. For metrics details see Section~\ref{subsec:evaluation_details}.}
\label{convergence}
\end{center}
\vskip -0.2in
\end{figure}

\begin{figure}[t]
\vskip 0.2in
\begin{center}
\centerline{\includegraphics[width=\columnwidth]{figures/LMperf_over_length.png}}
\caption{Full test set performance is better for later segments of sequences in keeping with intuition that additional context supports better predictions. We examined intervals up to $500$ tokens to ensure a minimum of $30$k samples per interval. }.
\label{LMperf_over_length}
\end{center}
\vskip -0.2in
\end{figure}

%% file: section/4-results.tex
\section{Results and Analysis}
\label{sec:results}

\begin{figure}[t]
\vskip 0.2in
\begin{center}
\centerline{\includegraphics[width=\columnwidth]{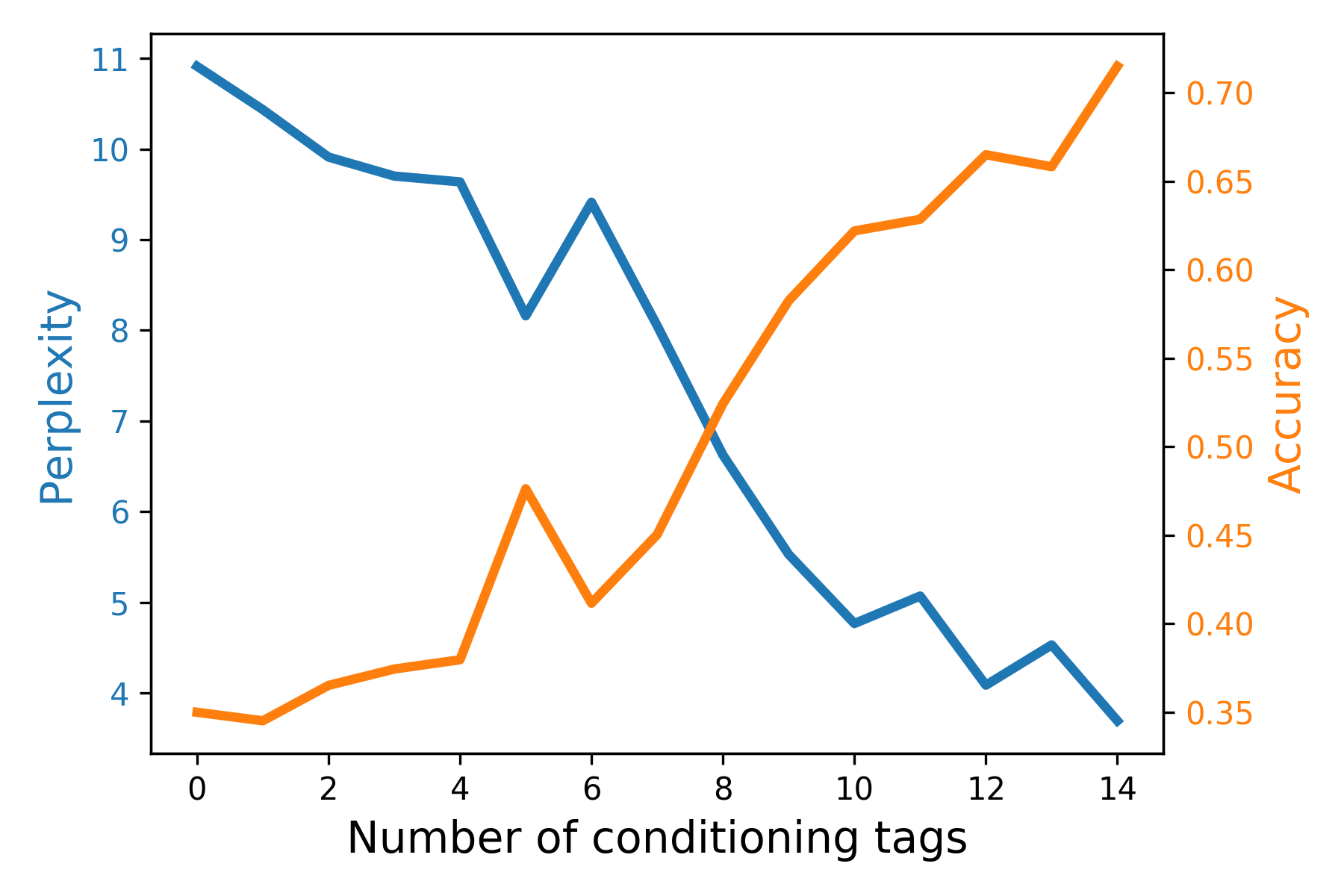}}
\caption{Full test set performance also improves as the number of conditioning tags associated with proteins increases. We examined proteins with up to $14$ conditioning tags to ensure a minimum of $3$k samples per category. }
\label{LMperf_over_codes}
\end{center}
\vskip -0.2in
\end{figure}
\label{subsec:perword}

\subsection{Evaluating ProGen as a language model}

In this section, we demonstrate that ProGen is a high-quality language model according to per-token metrics on the training and test sets. 

\textbf{ProGen generalizes to the full test set and achieves perplexities representative of a high-quality language model.}  
Perplexities reported in Table~\ref{test_set_ppl_acc} demonstrate that ProGen dramatically improves over a Uniform Baseline, in which amino acids are sampled according to a uniform distribution, and an Empirical Baseline, in which amino acids are sampled according to the empirical frequencies in the training set.
As a point of reference, state-of-the-art unidirectional language models for English Wikipedia achieve perplexities that range from $10$ to $17$ depending on model size (between $257$M and $8.3$B parameters) and whether training data was constrained to English Wikipedia~\citep{rae2019compressive} or not~\citep{shoeybi2019megatron}.

\textbf{ProGen generalizes to unseen protein families.}
The second section of Table~\ref{test_set_ppl_acc} breaks this result into perplexities over the ID-test and OOD-test sets separately.
Results on ID-test confirm that ProGen generalizes well to sequences that belonged to protein families randomly sampled.
As expected, performance is worse on the sequences in the OOD-test set, but the model still outperforms the Empirical Baseline for those held out protein families.

\textbf{Fine-tuning ProGen on unseen protein families improves over training from random initialization.}
We further split OOD-test into OOD-test-80 and OOD-test-20, fine-tuned ProGen on OOD-test-80 until convergence (5 epochs; Adam; linear learning rate warmup to 1k iterations), and retested on OOD-test-20. 
The third section of Table~\ref{test_set_ppl_acc} shows that fine-tuning from ProGen improves over training the same architecture with randomly initialized weights.
 
 \textbf{ProGen performance improves with increased amino acid and conditioning tag context.}
In Figure~\ref{LMperf_over_length}, we examine the mean perplexity and per-token hard accuracy over different portions of proteins. 
Perplexity decreases and hard accuracy increases for later portions of a protein, in keeping with the intuition that additional amino acid context narrows down the possibilities for future tokens.
The same trends hold when increasing the number of conditioning tags and taking the mean over  sequence lengths with the same of tags (in Figure~\ref{LMperf_over_codes}).
This indicates that conditioning tags also provide signal that improves  model predictions.

\textbf{Training curves suggest that protein generation would benefit from even larger models and longer training.}
With 1B parameters, ProGen is comparable in size to the largest language models that have been publicly released for any modality, and, to the best of our knowledge, it is the largest model trained on amino acid sequences. 
Figure~\ref{convergence} shows that despite its size and the amount of compute used to train, ProGen has yet to overfit the training data.
This suggests that models for protein generation could still benefit from even larger models and additional compute.

\textbf{BLOSUM62 soft accuracy reveals that ProGen prediction errors often follow natural amino acid substitutions that likely conserve higher level structure.}
Though ProGen models proteins as pure sequences, protein function is more directly determined by the secondary and tertiary structures that these sequences encode in three-dimensional space.
Model performance based on BLOSUM62 soft accuracy (Section~\ref{subsec:evaluation_details}) is more than 20\% higher than using hard accuracy, which indicates that when ProGen errors may often be substitutions that are acceptable in nature because they still reflect the proper higher-level properties.
This suggests that ProGen has learned how to work within function-preserving mutational invariances---we continue to validate this finding for primary, secondary, and conformational structure in Section~\ref{subsec:generation}.

\begin{figure}[t]
\vskip 0.2in
\begin{center}
\centerline{\includegraphics[width=\columnwidth]{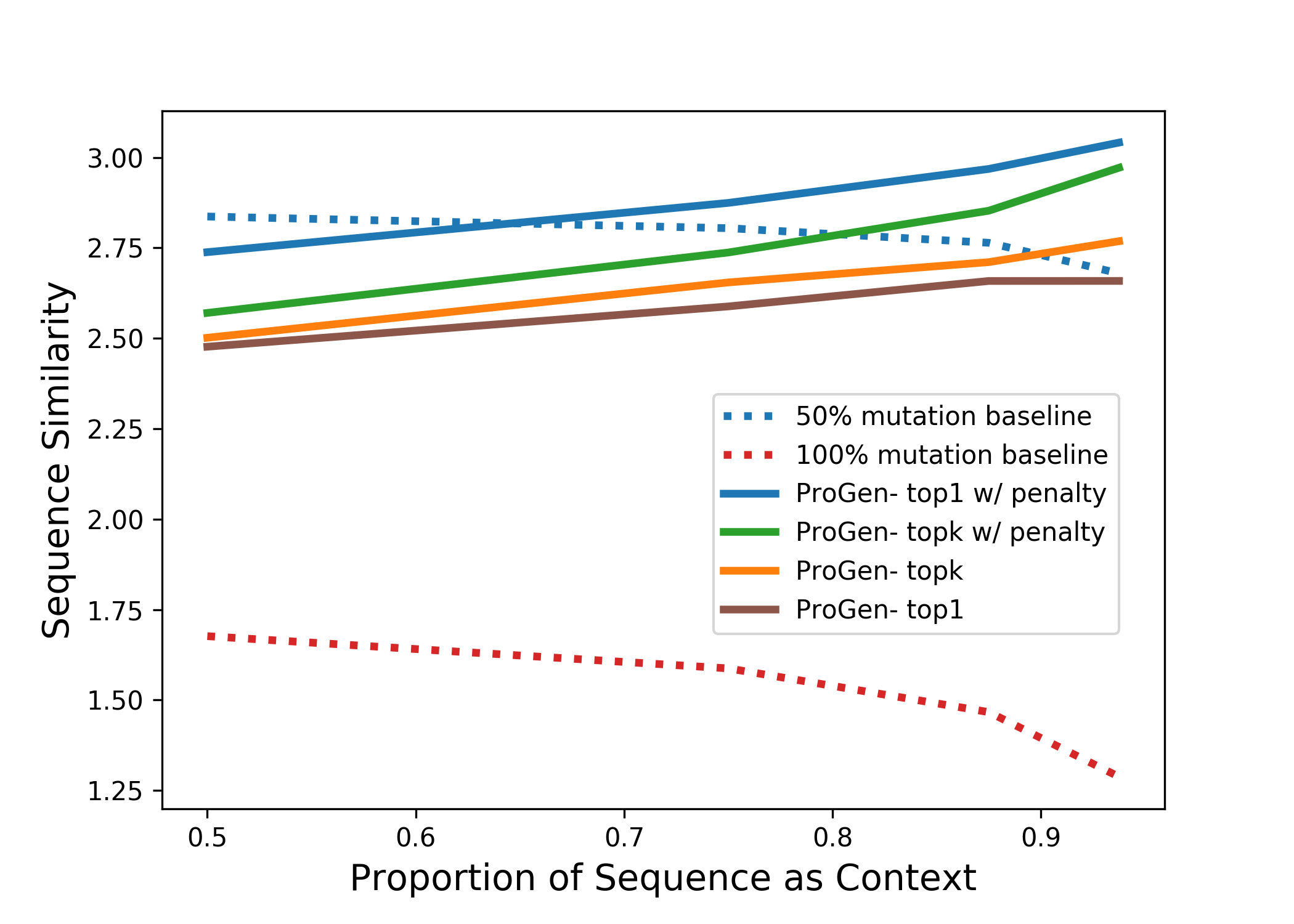}}
\caption{Across all context lengths, greedily sampling with a repetition penalty provides the best results according to sequence similarity. 
}
\label{seq_sim_base}
\end{center}
\vskip -0.2in
\end{figure}

\subsection{Generating with ProGen}
\label{subsec:generation}

In this section, we focus on assessing ProGen as a generative model.
Generation quality is directly correlated with evolutionary viability and functional qualities, which can be inferred through protein structure.
For this reason, we assess generation quality by using metrics for primary sequence similarity, secondary structure accuracy, and conformational energy (Section~\ref{subsec:evaluation_details}).

We also include several mutation baselines (Section~\ref{subsec:evaluation_details}) that allow us to compare the similarity of generated proteins to a target, reference protein across all metrics.
In reference to these mutation baselines, ProGen quality improves as we move from primary sequence to full conformational structure metrics, thereby suggesting the model has learned mutational invariances in structure which present as errors in lower-level metrics.

\textbf{ProGen achieves higher sequence similarity scores with an amino acid repetition penalty.}
Figure~\ref{seq_sim_base} depicts the results of experimenting with various combinations of top-$k$ sampling and repetition penalties (see  Section~\ref{subsec:evaluation_details} for details). 
Over all context lengths, ProGen performs best with $k=1$ and the repetition penalty applied to recently generated amino acids. 
Consequently, we use these settings for all following generation experiments. With this nearly greedy sampling, ProGen manages to generate proteins with sequence similarity comparable to randomly mutating $50$\% of the amino acids that are not seen in the given context. 

\begin{figure}[h!]
\vskip 0.2in
\begin{center}
\centerline{\includegraphics[width=\columnwidth]{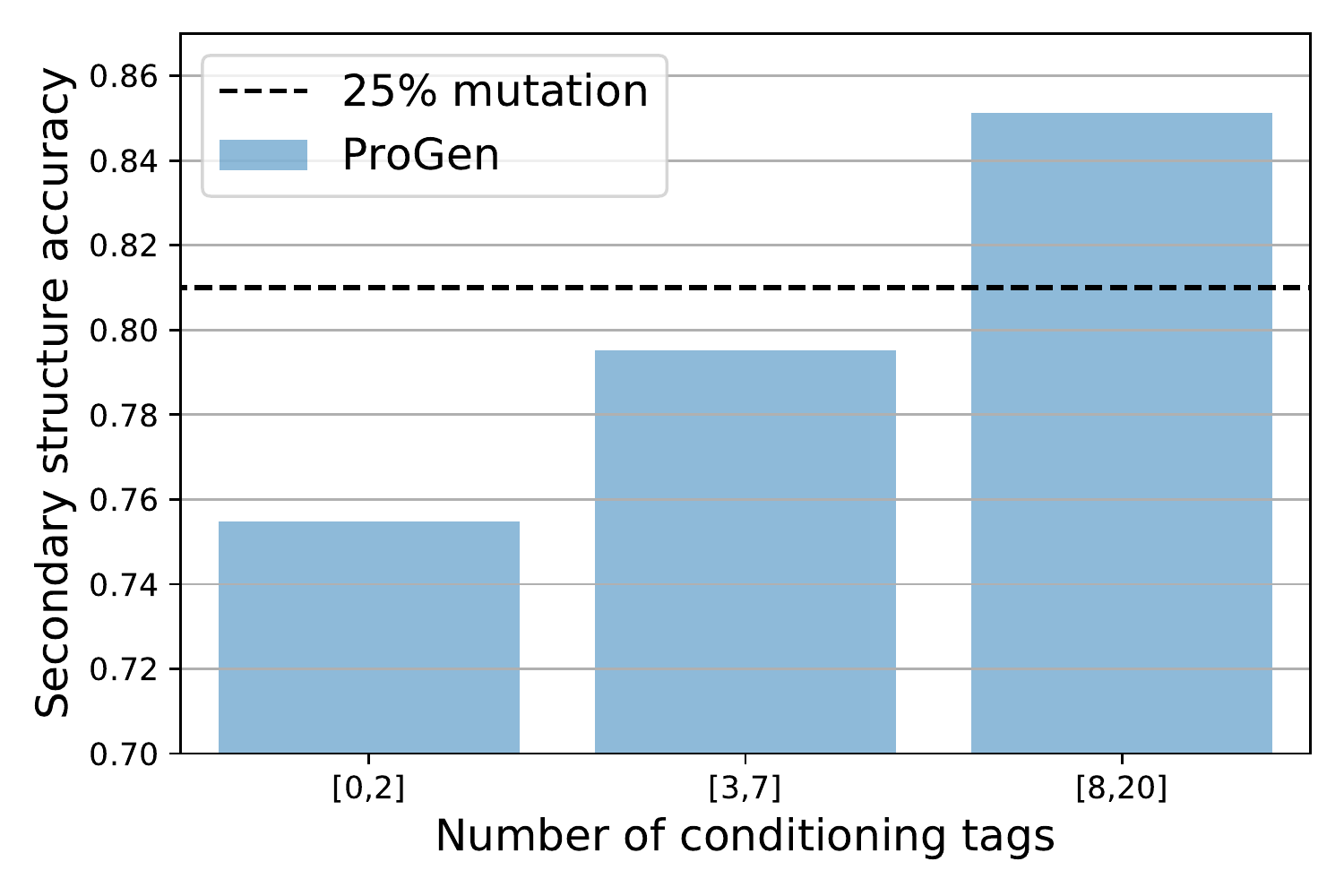}}
\caption{ProGen generates sequences that conserve secondary structure of the protein. Increasing the number of conditioning tags yields better secondary structure accuracy than the 25\% mutation baseline.} 
\label{ss_align}
\end{center}
\vskip -0.2in
\end{figure}

\textbf{Sequence similarity suggests that ProGen merely approaches the 25\% mutation baseline, but secondary structure accuracy suggests that ProGen surpasses it.} 
\begin{figure}[t]
\vskip 0.2in
\begin{center}
\centerline{\includegraphics[width=1\columnwidth]{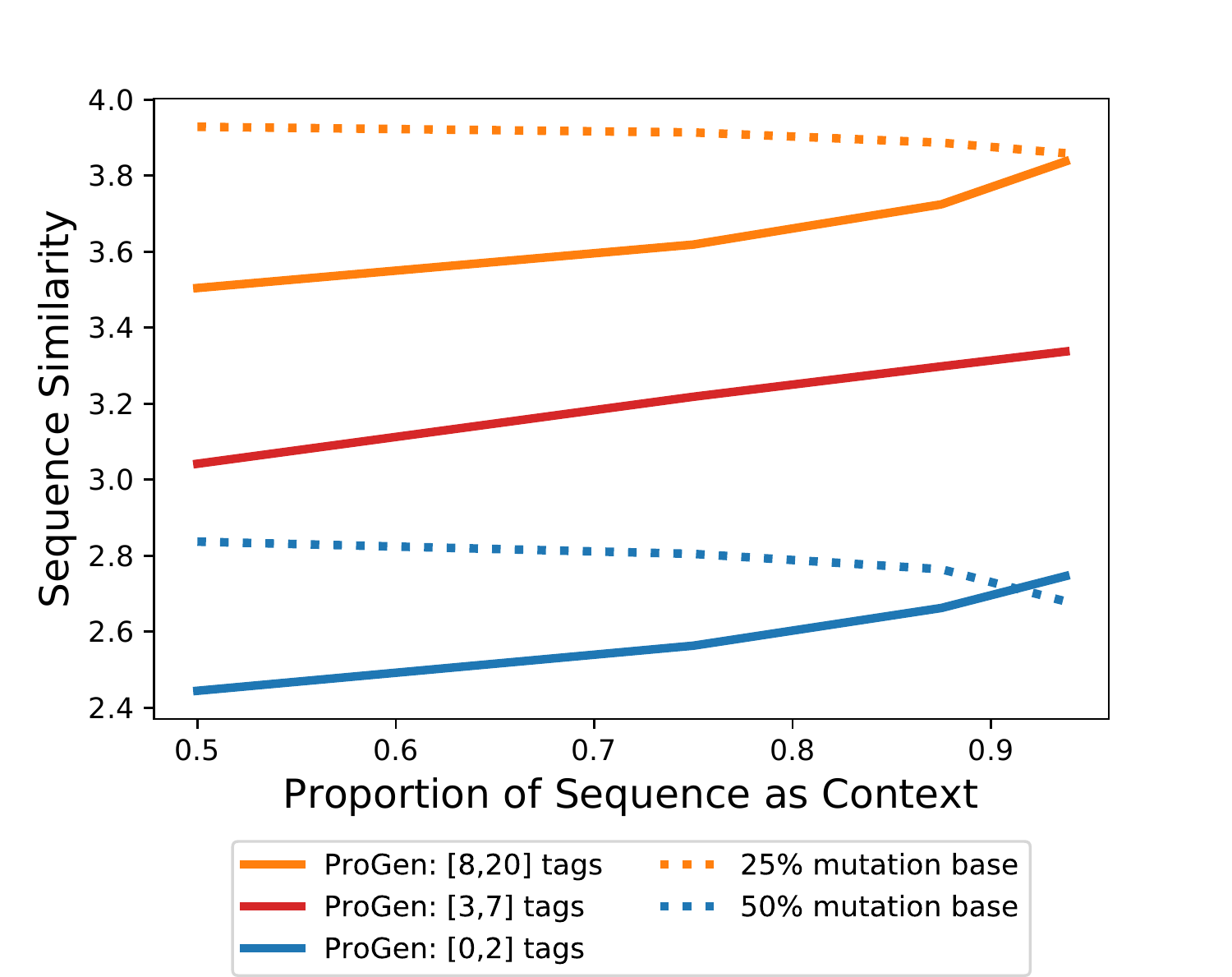}}
\caption{A greater number of conditioning tags enables higher quality generation. With at least $8$ conditioning tags, generation quality approaches the 25\% mutation baseline.
}
\label{seq_align_kws}
\end{center}
\vskip -0.2in
\end{figure}
In Figure~\ref{seq_align_kws}, we analyze this sequence similarity across differing numbers of conditioning tags. 
Sequences associated with at least $3$ conditioning tags begin to exceed the $50$\%  mutation baseline, and as amino acid context increases, sequences with at least $8$ conditioning tags approach the $25$\% mutation baseline.
Notably, even in the best case, according to sequence similarity, ProGen doesn't surpass the $25$\% mutation baseline.
By contrast, according to secondary structure accuracy, sequences with at least $8$ conditioning tags surpass the $25$\% mutation baseline (Figure~\ref{ss_align}).
This discrepancy between sequence similarity and secondary structure accuracy further corroborates our claim from Section~\ref{subsec:perword}:
errors registered by lower-level metrics often correspond to acceptable substitutions according to higher-level metrics that more directly correspond to functional viability.

\begin{figure}[h]
\vskip 0.2in
\begin{center}
\centerline{\includegraphics[width=\columnwidth]{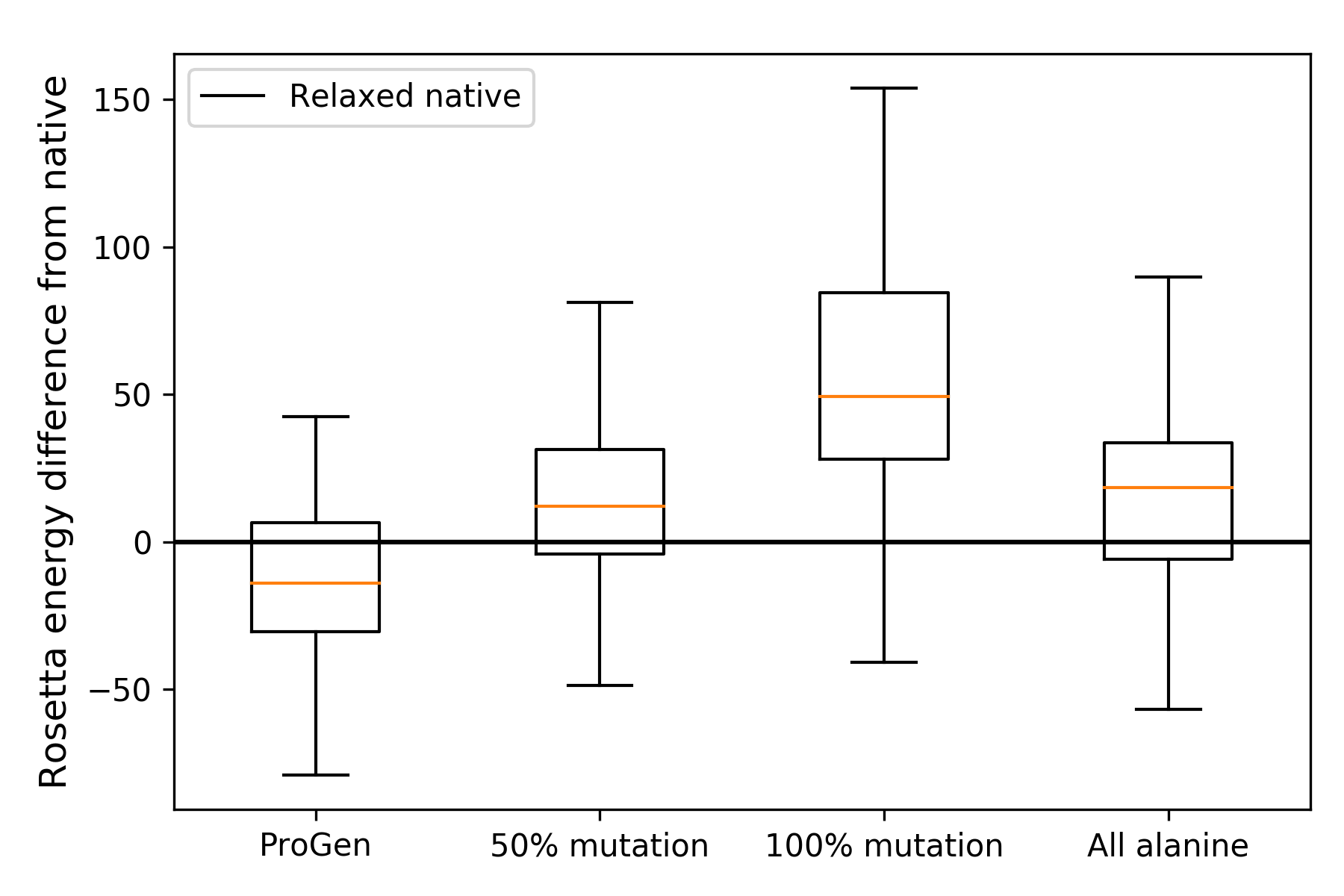}}
\caption{Conformational energies for ProGen generated proteins surpasses all baselines and adheres closely to the energy of the native template.}
\label{test_structures}
\end{center}
\vskip -0.2in
\end{figure}

\textbf{After threading and relaxation, samples generated by ProGen are likely to exhibit desired structure and function.}
As a measure of generation quality, we thread ProGen sequences through known structures and examine if they exhibit favorable, low energy states. Figure~\ref{test_structures} shows the differences between the energy levels of native proteins, ProGen samples, the native proteins with $50$\% and $100$\% of amino acids randomly mutated, as well as the all-alanine baseline. 
Proteins completed by ProGen are much closer to the energy levels of the native protein than all baselines.
Generated samples exhibit energy levels near or even below their associated relaxed native templates. 

\begin{figure}[h!]
\vskip 0.2in
\begin{center}
\centerline{\includegraphics[width=\columnwidth]{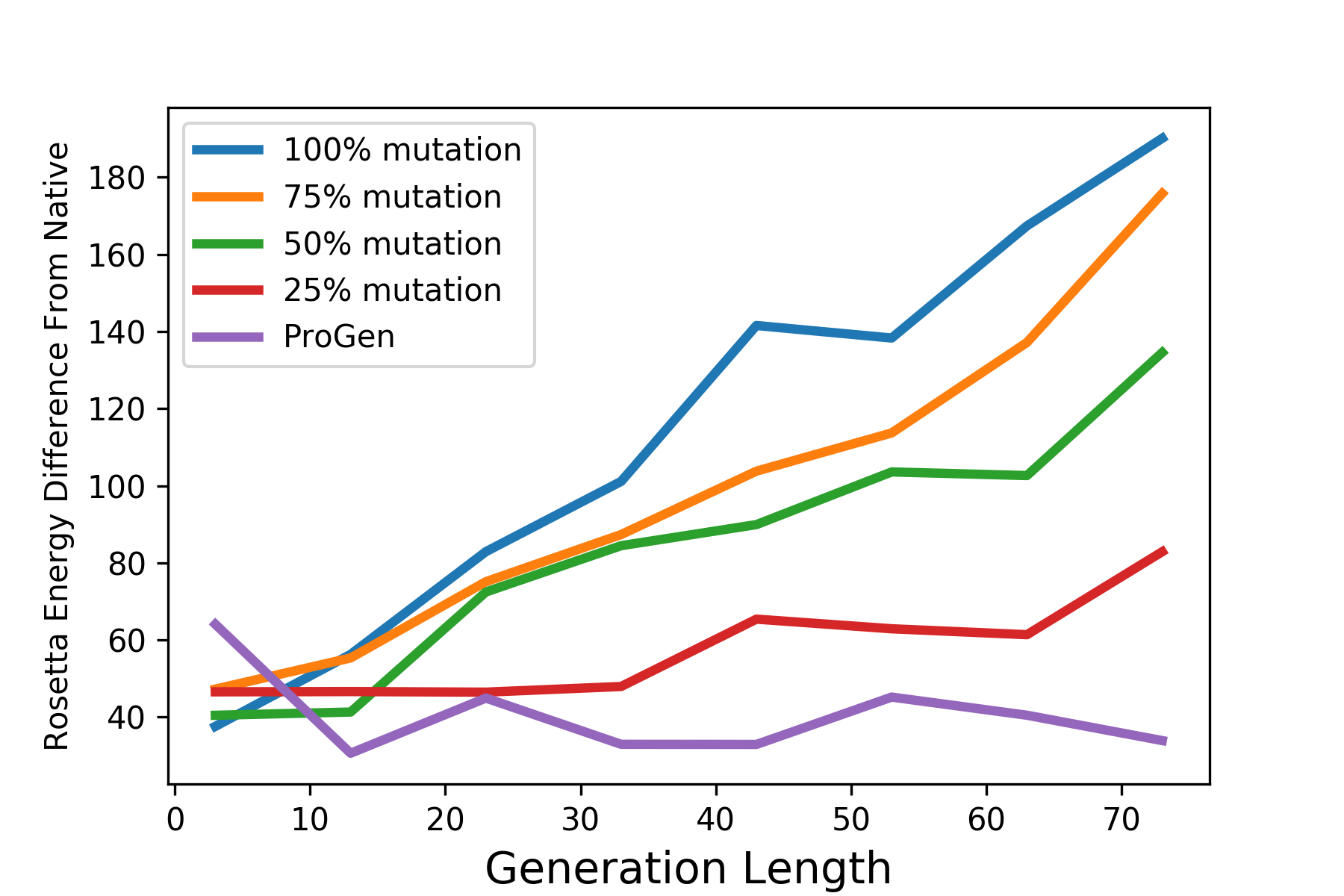}}
\caption{ProGen completion quality for VEGFR2 remains steadily near native conformational energy levels across generation lengths.}
\label{vegfr2_energy}
\end{center}
\vskip -0.2in
\end{figure}

\subsection{Case Study: Completing VEGFR2 kinase domain}
\begin{figure}[h]
\vskip 0.2in
\begin{center}
\centerline{\includegraphics[width=\columnwidth]{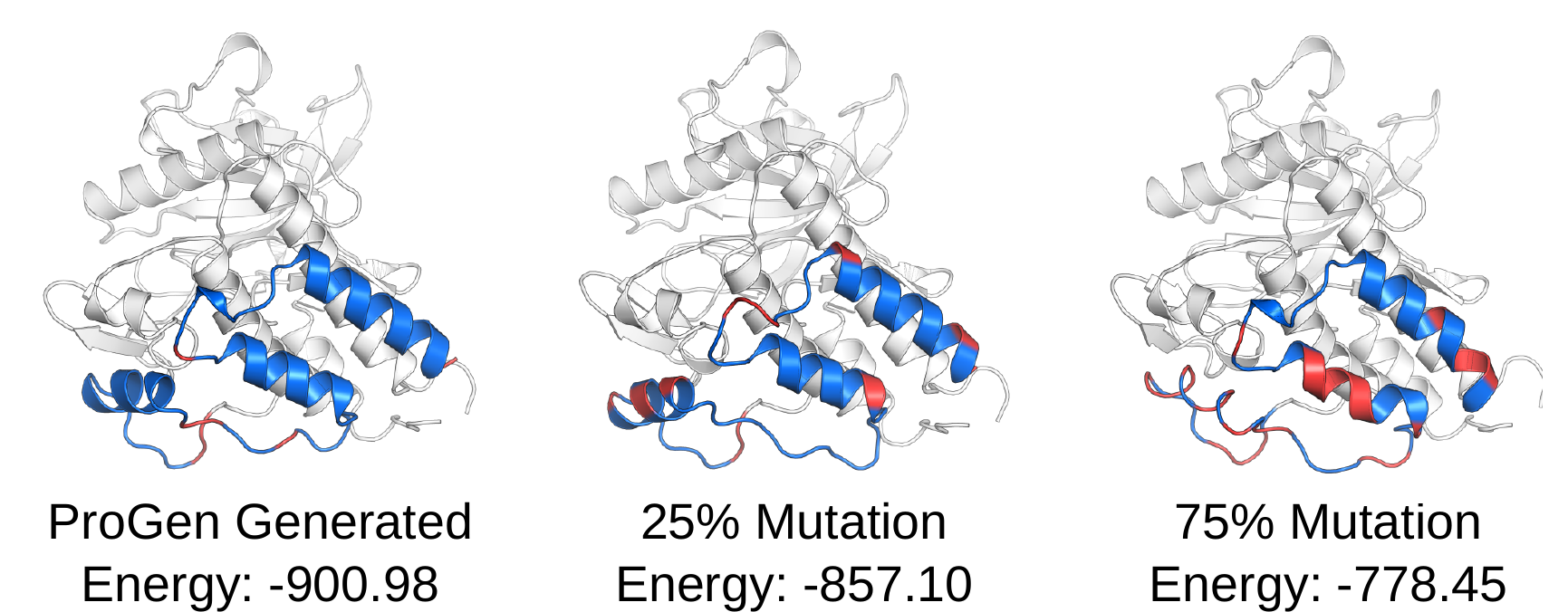}}
\caption{ ProGen makes fewer mistakes and prioritizes conservation of secondary structure as compared to baselines. Blue is low energy (stable) and red high (unstable).} 
\label{vegfr2_samples}
\end{center}
\vskip -0.2in
\end{figure}
VEGFR2 is responsible for fundamental cell processes such as cell proliferation, survival, migration, and differentiation. VEGFR2 was excluded from training as a subsequence belongs to a held out protein family in OOD-test. We study how well ProGen generates in the context of a protein completion task. 
We consider the amino acid sequence beginning at residue 806 and ending at residue 1168 of VEGFR2 (PDB ID: 2XIR). 
For different generation lengths, we sample from ProGen to complete the sequence up to residue 1168 with the remainder of the sequence provided as context. 
Figure~\ref{vegfr2_energy} shows that the conformational energy calculated after threading and relaxation of ProGen samples are lower compared to all baselines, indicating better structural conservation. 
Generation quality remains near the native relaxed protein independent of generation length. 

The generated samples across Figure~\ref{vegfr2_energy} exhibit a mean sequence identity of $73.1\%$ with the native sequence. 
This correlates to a lower sequence identity than the $25\%$ mutation baseline ($74\%$ identity) but with better Rosetta energies.
This suggests meaningful deviation from the native protein while achieving the ultimate goal of preserving low energy.

Figure~\ref{vegfr2_samples} shows one sample from ProGen as well as one from each of the 25\% and 75\% mutation baselines.
The ProGen sample exhibits lower energy overall, and energy is highest for amino acids that do not have secondary structure.
This suggests that ProGen learned to prioritize the most structurally important segments of the protein.

\subsection{Case Study: Zero-shot fitness selection for protein GB1}

The ultimate goal of protein engineering is to engineer \textit{functional} proteins. One promising avenue is via directed evolution, which iterates through rounds of mutation and screening to converge on a high-fitness (i.e. functioning) protein. 
Machine learning has shown initial promise to aid in the subsequent rounds of directed evolution by \textit{in silico} screening of proteins \citep{wu2019machine}, but it still relies on random mutation in an exponentially large search space. Ideally, a generative model, such as ProGen, that has learned the distribution of evolutionarily-relevant proteins can directly generate high-fitness proteins.

We examine the empirical fitness landscape of protein G domain B1 (GB1) binding to an antibody \citep{wu2016adaptation}.
Protein G is important for the purification, immobilization, and detection of immunoglobulins (antibodies), proteins used by our immune system to neutralize pathogenic viruses and bacteria. Ideally, we would want the ability to generate GB1 proteins with high binding affinity and stability.
The data includes 149,361 of a total 160,000 possible variants from NNK/NNS saturation mutagenesis at four positions known to interact epistatically. Reported fitness values correspond to a measure of both stability (i.e. the fraction of folded proteins) and function (i.e. binding affinity to IgG-Fc) by coupling mRNA display with next-generation sequencing.
Protein sequences with high fitness values are desired.
\begin{figure}[h!]
\vskip 0.2in
\begin{center}
\centerline{\includegraphics[width=\columnwidth]{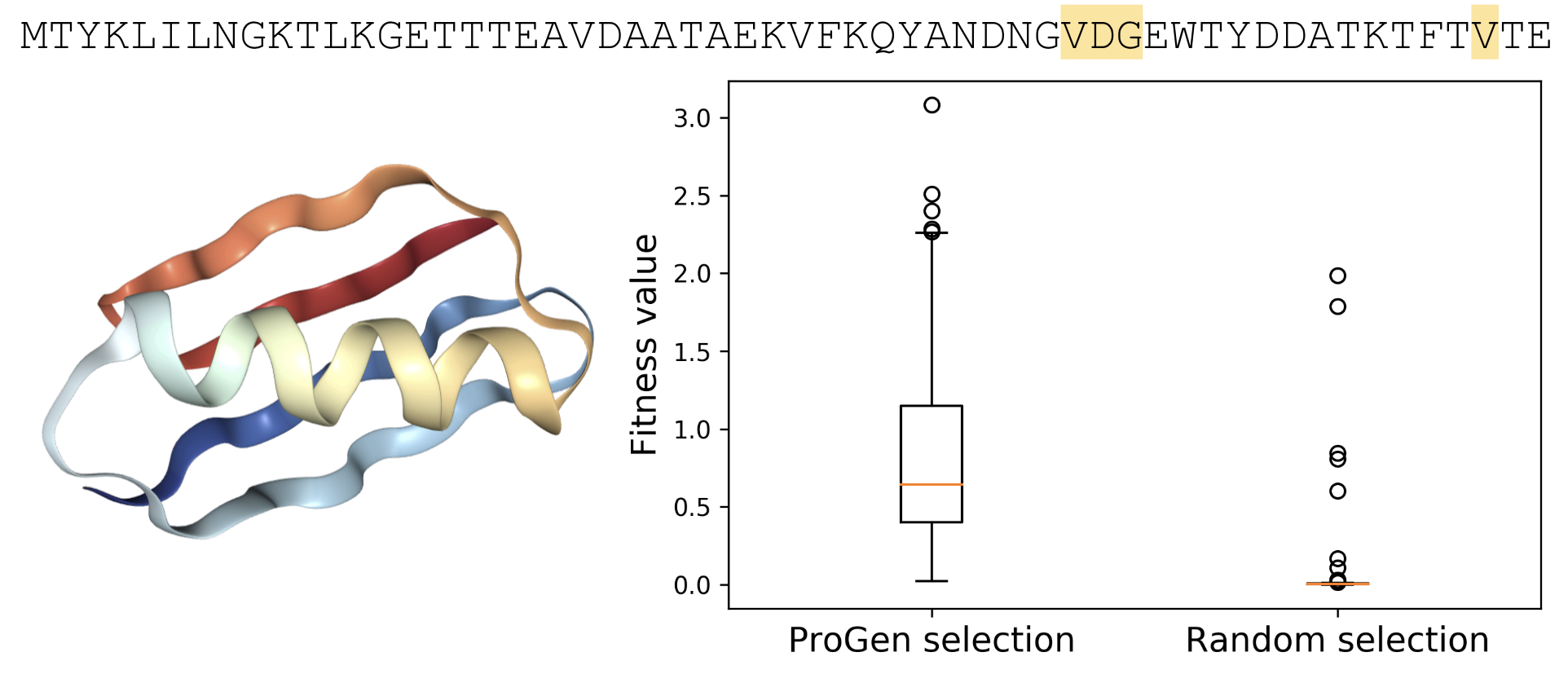}}
\caption{Without training on the \citet{wu2016adaptation} dataset, ProGen can identify which protein variants exhibit high fitness. The dataset reports fitness values for protein variants of GB1 binding to an antibody. Each sample corresponds to mutating one of four highlighted residues, in the above sequence, to a standard amino acid. At the left, the crystallized structure of GB1 is shown. At the right, the fitness value of samples selected through ProGen vs random selection are shown.} 
\label{GB1}
\end{center}
\vskip -0.2in
\end{figure}

Without supervised training of ProGen on the GB1 data or unsupervised fine-tuning of ProGen on a subset of similar immunoglobulin-binding proteins, we pass each variant through ProGen and select the top one hundred variants with the lowest perplexity values. In Figure \ref{GB1}, we demonstrate ProGen is effective in zero-shot selection of high-fitness protein sequences. In comparison, random mutation, which is the main technique used by directed evolution and ML-assisted directed evolution, statistically generates samples with low or zero fitness. With effective sampling techniques, ProGen can be utilized to generate a spread of samples that are statistically high fitness. These results imply that ProGen has not only learned the distribution of structurally-relevant proteins, but also functionally-relevant proteins.

%% file: section/5-conclusion.tex
\section{Conclusion}
\label{sec:conclusion}
We introduced ProGen, a controllable protein generation language model trained on the full evolutionary diversity of one of the largest sequence databases. 
The model generates proteins that exhibit near native structure energies which likely implies functional viability. 
ProGen has the potential to play a new, complementary role alongside other state-of-the-art methods in protein engineering.
For example, in directed evolution, initial sequences may be sampled from ProGen according to desired conditioning tags.
In later rounds of evolution, protein completion with context for particular residue spans, or hotspots, may provide higher fitness samples.
In {\it de novo} protein design, using ProGen with conditioning tags may allow for designing new proteins with existing folding motifs in new protein families or host organisms.
This same strategy may be used in conjunction with threading and structure-based protein design.
Because conditioning tags orient ProGen in sequence space, ProGen may even be used as a model to sample from the distribution of evolutionarily viable proteins near one particular protein.
This may provide useful augmentations around data for non-homologous domains where existing techniques, such as MSAs, fall short.

%% file: section/6-acknowledgements.tex
\section{Acknowledgements}
\label{sec:acknowledgements}
We would like to thank Alex Chu for assistance in the threading and minimization experiments along with Jesse Vig for visualizing the attention heads of ProGen.

%% file: section/8-appendix.tex
\section{Appendix}
\subsection{Measuring out-of-distribution}
\label{sec:OOD}

The objective of our work is to enable high-quality protein generation. To test the effectiveness of our trained model, we had two test subsets: ID-Test and OOD-Test.
ID-Test is a random split of the non-redundant sample database and can be viewed as a typical \textit{in-distribution} test set of held-out samples. 

In contrast, OOD-Test represents an \textit{out-of-distribution} set. 
OOD-Test consists samples that contained a matching sub-sequence residing in one of twenty Pfam protein families that were held out of Train and ID-Test.

\begin{table}[h]
\vskip 0.15in
\begin{center}
\begin{small}
\begin{sc}
\begin{tabular}{lccr}
\toprule
                    & 3-gram SAE & 5-gram SAE \\
\midrule
Train and ID-Test   & 0.027      & 0.095      \\
Train and OOD-Test   & 0.399      & 1.112      \\
ID-Test and OOD-Test & 0.387      & 1.104     \\
\bottomrule
\end{tabular}
\end{sc}
\end{small}
\end{center}
\caption{The training data and ID-Test data seem to be drawn from a similar distribution, but OOD-Test is markedly different from the others. SAE refers to the sum of absolute errors for normalized 3-gram and 5-gram histograms. If two histograms were entirely divergent, the SAE would yield a value of 2.}
\label{ngram}
\end{table}

To quantify the out-of-distribution nature of OOD-Test, we computed a normalized histogram of 3-grams and 5-grams across samples in the Train, ID-Test, and OOD-Test datasets. The sum of absolute errors (SAE) was computed for a pair of histograms as shown in Table~\ref{ngram}. Two normalized histograms that align perfectly would have an SAE of 0 and two normalized histograms that are completely divergent would have an SAE of 2. The results imply that the OOD-Test is drawn from a significantly different distribution.

The held-out protein families included \texttt{PF18369, PF04680, PF17988, PF12325, PF03272, PF03938, PF17724, PF10696, PF11968, PF04153, PF06173, PF12378, PF04420, PF10841, PF06917, PF03492, PF06905, PF15340, PF17055, PF05318}.

\subsection{Generation with only conditioning tags}
We observe that ProGen can be used to generate proteins with only conditioning tags and no initial amino acid context. For the following example, we prompt ProGen to greedily generate a protein sequence with the tags \texttt{Flavoprotein} and \texttt{FMN}. As defined by the UniprotKB keyword, the \texttt{FMN} tag refers to ``a protein involved in flavin adenine mononucleotide (FMN) synthesis or protein which contains at least one FMN as prosthetic group/cofactor (flavoproteins) or cosubstrate, such as many oxidation-reduction enzymes''.

\begin{figure*}[hbtp]
\vskip 0.2in
\begin{center}
\centerline{\includegraphics[width=\textwidth]{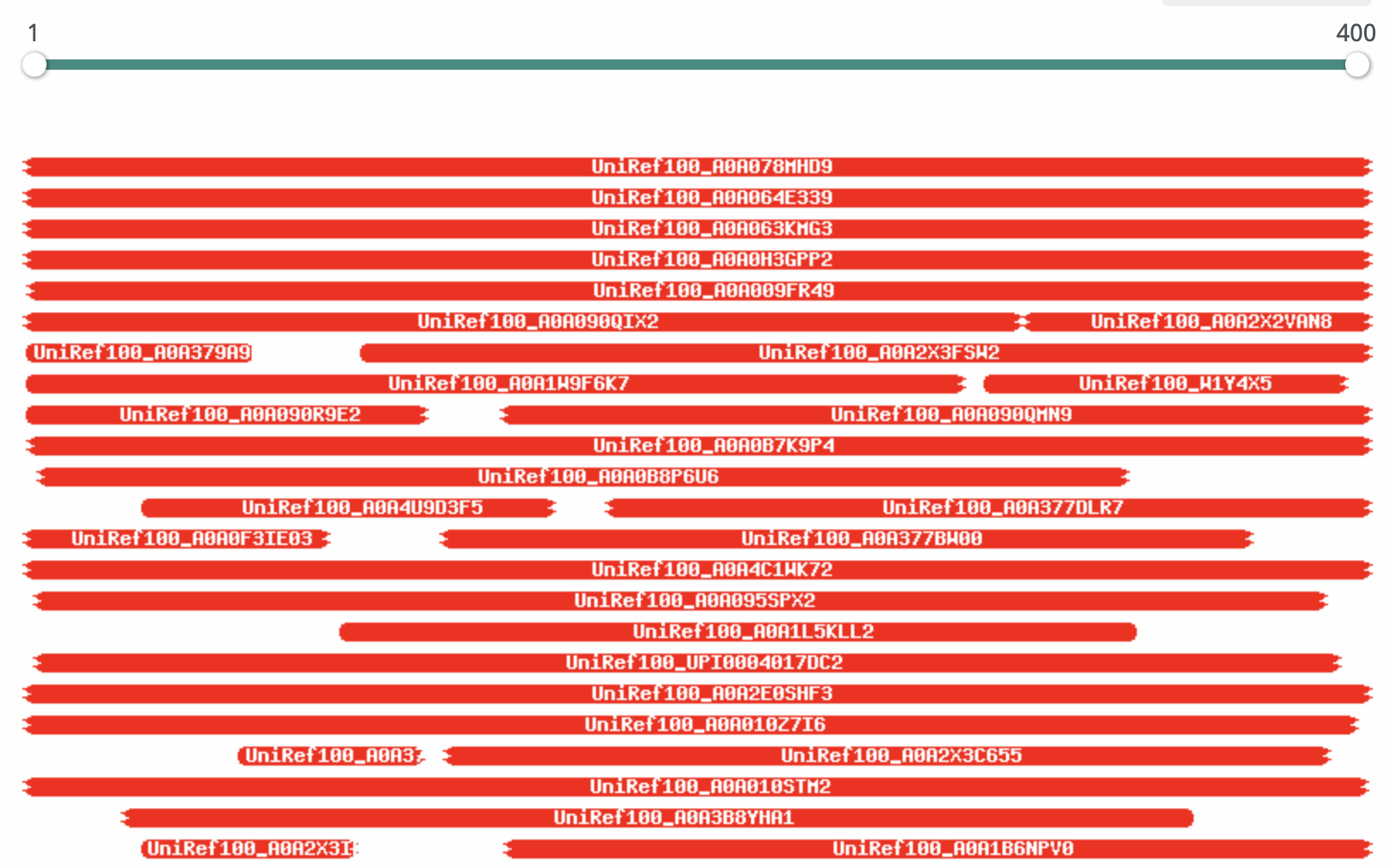}}
\caption{There are multiple sequences that align well with the ProGen generated FMN sequence from only conditioning tags. Many of the matching alignments have properties reflective of FMN proteins (e.g. oxidoreductases). A red color corresponds to a significantly low E-value, implying a matching homolog. The MSA was directly taken using HHblits.}
\label{FMN_overlap}
\end{center}
\vskip -0.2in
\end{figure*}

\begin{figure*}[hbtp]
\vskip 0.2in
\begin{center}
\centerline{\includegraphics[width=\textwidth]{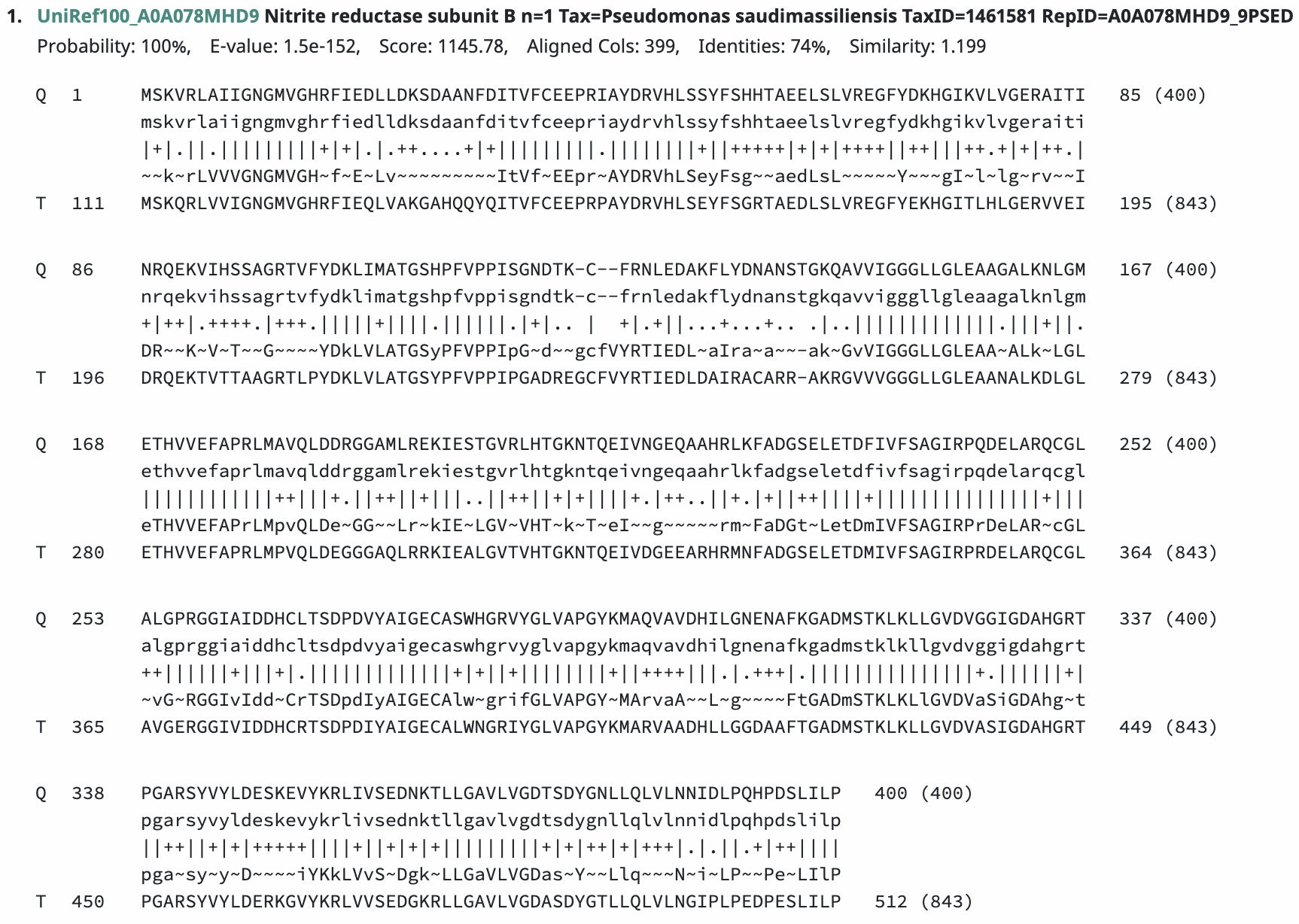}}
\caption{First alignment (ranked by E-value) of a ProGen generated FMN protein. An E-value less than $1e^{-4}$ and identity greater than $40\%$ is desired to consider the match as potentially homologous. The sequence labeled as Q is the ProGen protein and the sequence labeled as T is the matched sequence.}
\label{FMN_1}
\end{center}
\vskip -0.2in
\end{figure*}

\begin{figure*}[hbtp]
\vskip 0.2in
\begin{center}
\centerline{\includegraphics[width=.85\textwidth]{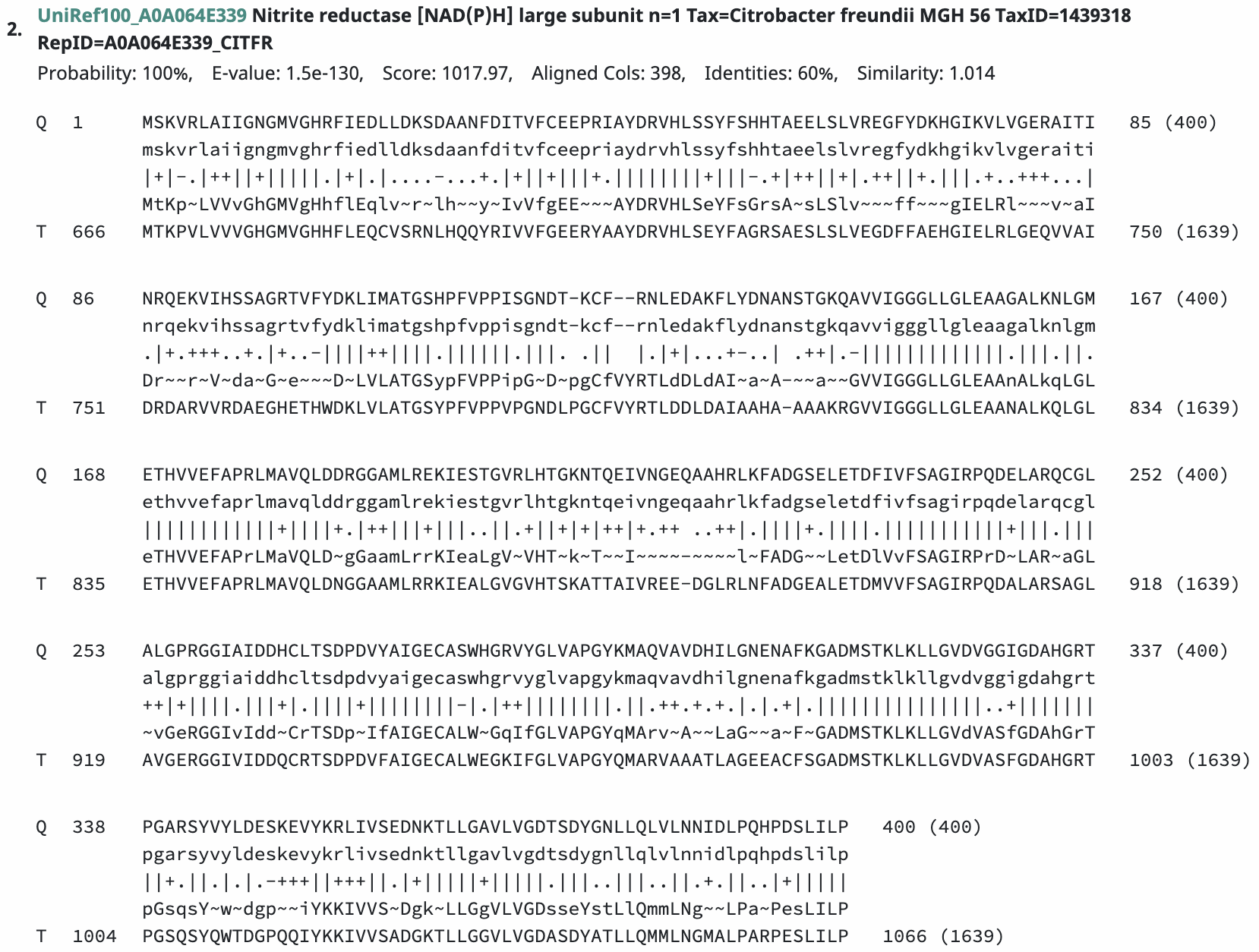}}
\caption{Second alignment (ranked by E-value) of a ProGen generated FMN protein. An E-value less than $1e^{-4}$ and identity greater than $40\%$ is desired to consider the match as potentially homologous. The sequence labeled as Q is the ProGen protein and the sequence labeled as T is the matched sequence.}
\label{FMN_2}
\end{center}
\vskip -0.2in
\end{figure*}

\begin{figure*}[hbtp]
\vskip 0.2in
\begin{center}
\centerline{\includegraphics[width=.85\textwidth]{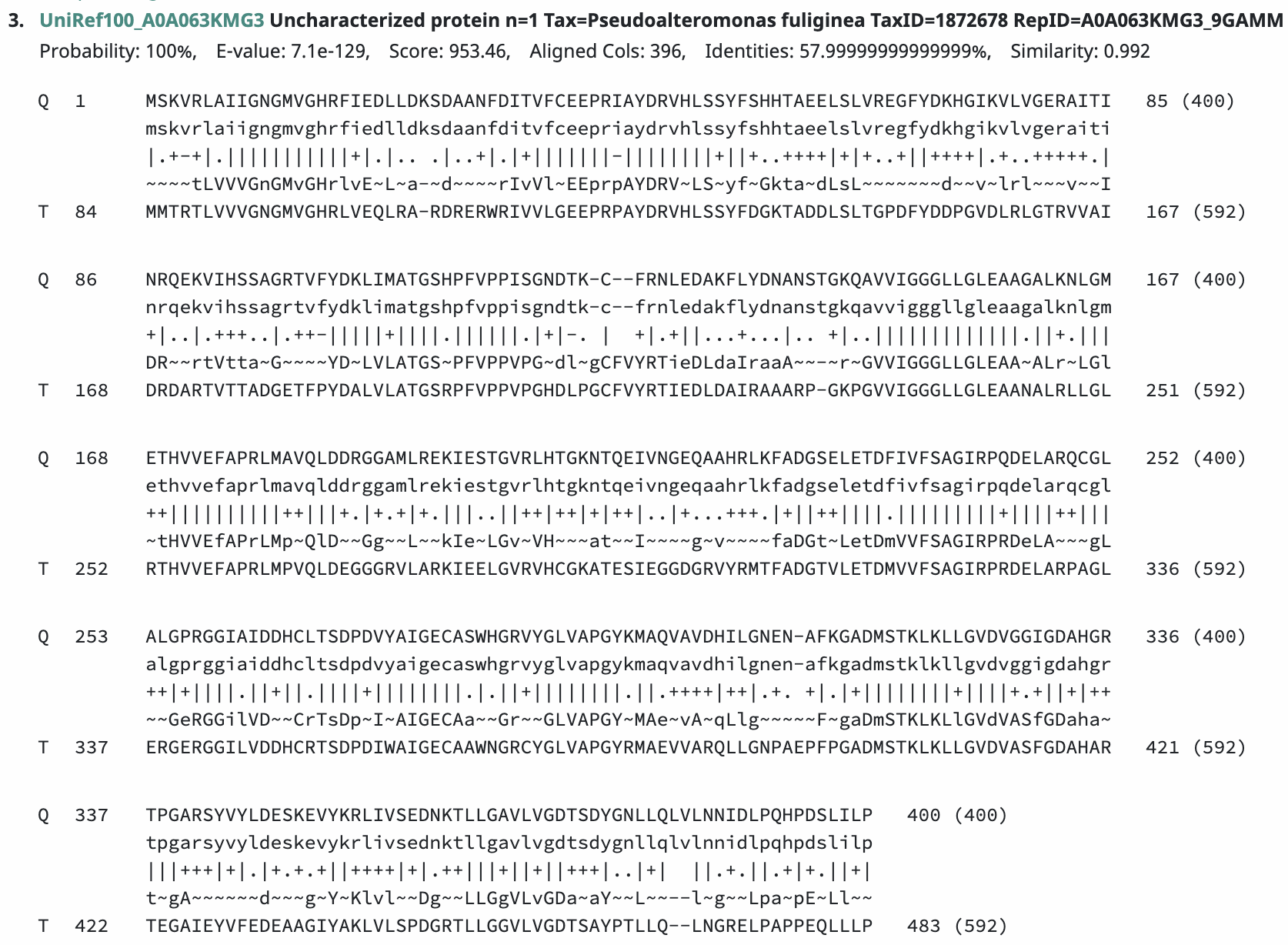}}
\caption{Third alignment (ranked by E-value) of a ProGen generated FMN protein. An E-value less than $1e^{-4}$ and identity greater than $40\%$ is desired to consider the match as potentially homologous. The sequence labeled as Q is the ProGen protein and the sequence labeled as T is the matched sequence.}
\label{FMN_3}
\end{center}
\vskip -0.2in
\end{figure*}

The generated sequence of length 400 is then passed to the HHblits package by \citet{zimmermann2018completely} to search for a multiple sequence alignment (MSA). As shown in Figure~\ref{FMN_overlap}, there are multiple sequences that align well with the ProGen sequence. Figures ~\ref{FMN_1}-\ref{FMN_3} demonstrate the alignments have high E-values and have related properties. The lower the E-value, the lower the probability of a random match and the higher the probability that the alignment match is related to the searched sequence.

\subsection{Model visualizations}
\begin{figure}[h]
\vskip 0.2in
\begin{center}
\centerline{\includegraphics[width=\columnwidth]{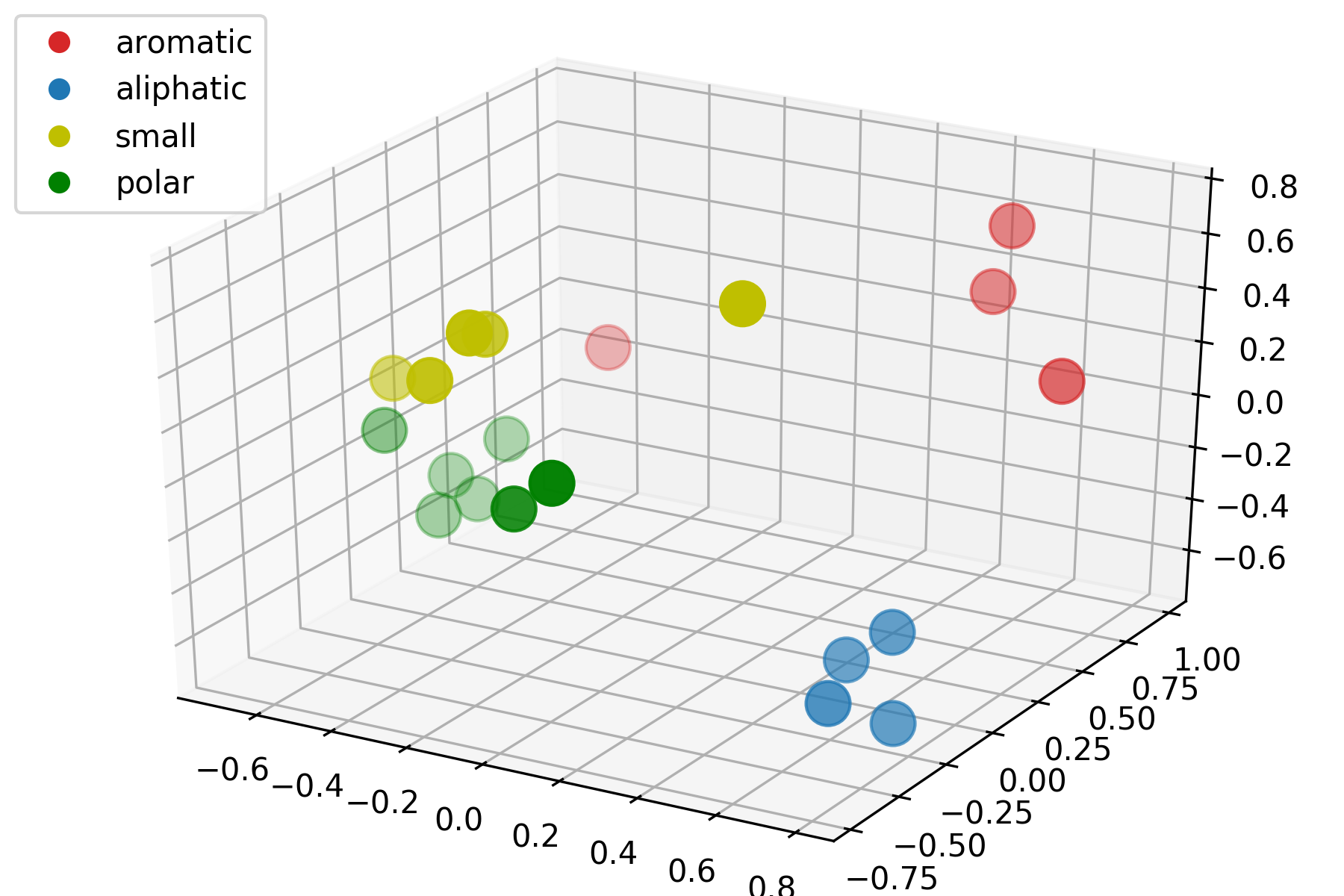}}
\caption{Principle component analysis (PCA) of the ProGen's amino acid embeddings aligns with our intuition of amino acid properties.}
\label{pca}
\end{center}
\vskip -0.2in
\end{figure}

ProGen was trained from a randomly initialized embedding layer with no prior knowledge of residue biochemical properties. Through per-token training on millions of protein sequences, ProGen seems to have inherently learned the natural clustering of amino acids that align with our understanding of biophysicochemical properties. In Figure~\ref{pca}, the trained embedding weights for the standard amino acids tokens are reduced to three dimensions with principle component analysis (PCA).

\begin{figure*}[h!]
\vskip 0.2in
\begin{center}
\centerline{\includegraphics[width=.9\textwidth]{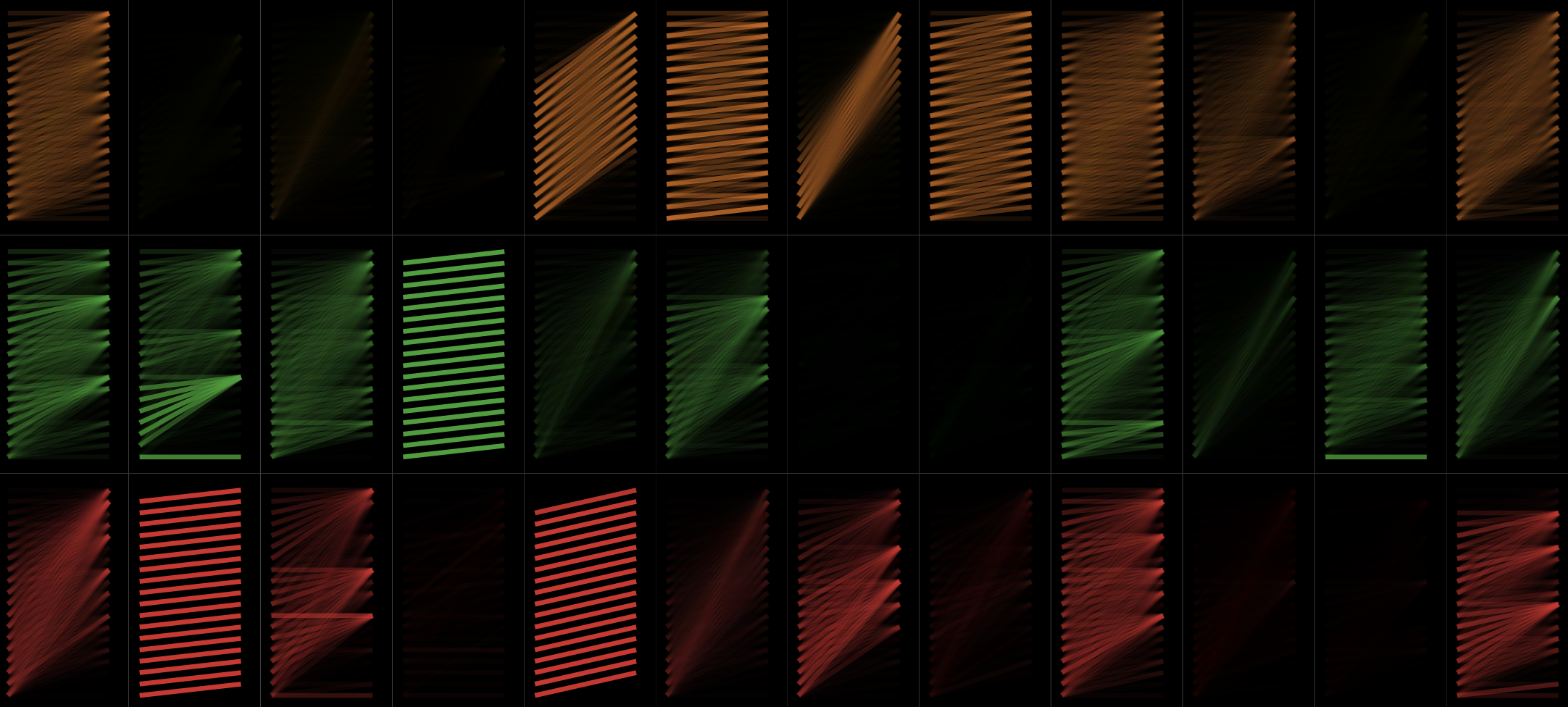}}
\caption{Attention patterns of ProGen for a given sequence. Layers 1-3 (rows) and attention heads 1-12 (columns) are displayed. The attention mechanism exhibits well-differentiated local and global patterns which may indicate specialization of each head on different tasks. Two corresponding attention heads from this visualization are shown in Figure~\ref{att_heads_2}.}
\label{all_att_heads}
\end{center}
\vskip -0.2in
\end{figure*}

\begin{figure*}[hbtp]
\vskip 0.2in
\begin{center}
\centerline{\includegraphics[width=.9\textwidth]{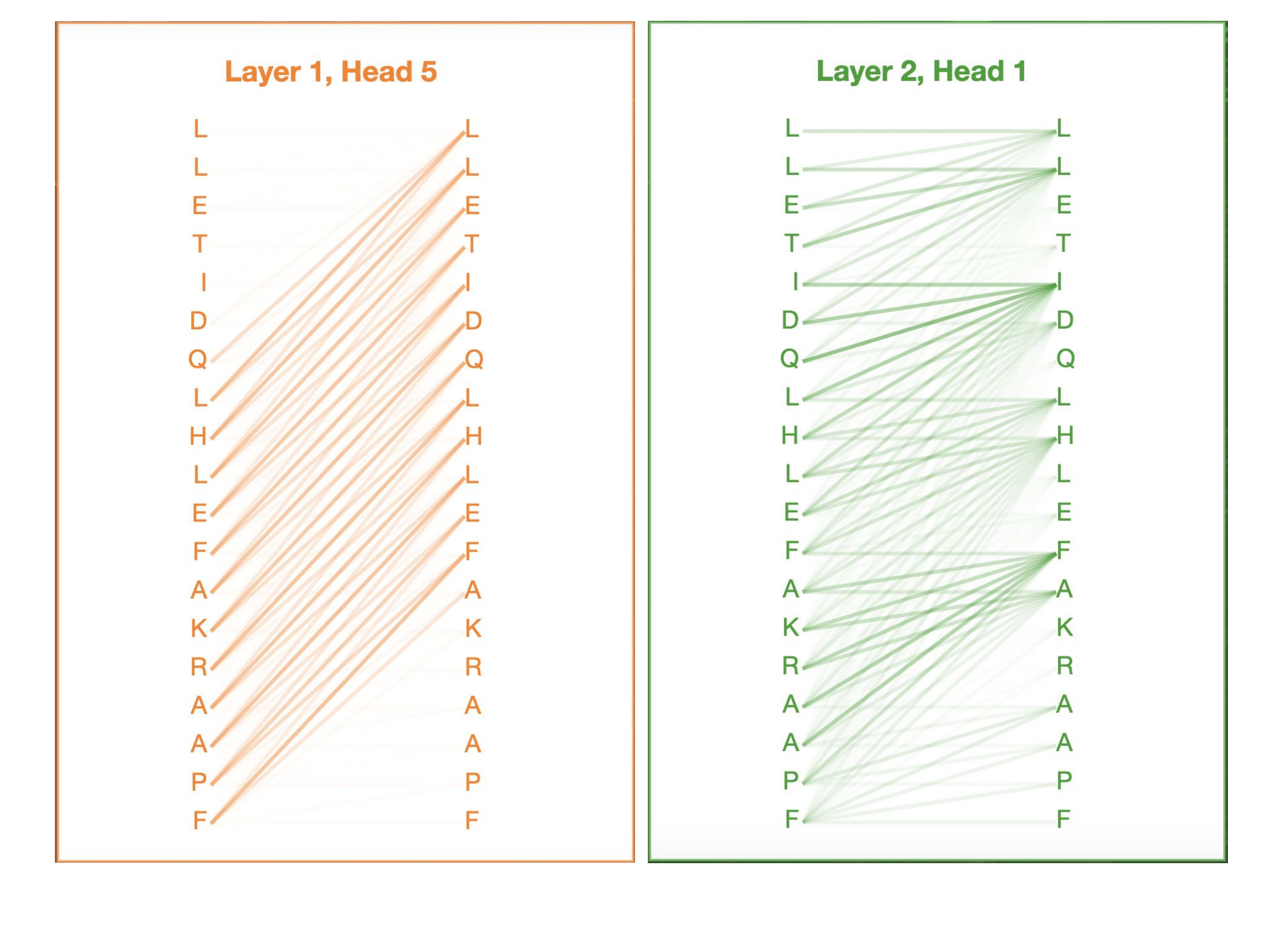}}
\caption{Local attention pattern for two example attention heads. Lines indicate attention to previous tokens for a given predicted token.}
\label{att_heads_2}
\end{center}
\vskip -0.2in
\end{figure*}

Using \citet{vig2019multiscale}, we visualize the attention head patterns of ProGen. For both Figure ~\ref{all_att_heads} and Figure~\ref{att_heads_2}, we are visualizing the attention weight patterns in each head of ProGen for $\alpha$-actinin protein (PDB: 4D1E) residues 510 to 528, which exhibits an alpha helical structure. In Figure~\ref{all_att_heads}, we visualize layers 1 to 3 and attention heads 1 to 12 of ProGen. The attention mechanism exhibits well-differentiated local and global patterns which may indicate specialization of each head on different tasks.

